\documentclass{an}
\usepackage{graphicx}
\usepackage{times}
\usepackage{fancyhdr}
\begin{document}

\title{A possible interrelation between the estimated luminosity
distances and internal extinctions of type Ia supernovae}

\author{L.G. Bal\'azs \inst{1}
\and
 Zs. Hetesi \inst{2}
 \and
Zs. Reg\'aly \inst{1} \inst{2} \and Sz. Csizmadia \inst{1} \and
Zs. Bagoly \inst{3} \and  I. Horv\'ath \inst{4} \and A.
M\'esz\'aros \inst{5}}

\institute{Konkoly Observatory, Budapest, Box-67, H-1525, Hungary\\
    \email{balazs@konkoly.hu, csizmadi@konkoly.hu}
    \and Department of Astronomy, E\"otv\"os University, Budapest,
    Budapest, P\'azm\'any P. s. 1/A,, H-1117,
          Hungary\\
          \email{Zs.Hetesi@astro.elte.hu, Zs.Regaly@astro.elte.hu}
\and Laboratory for Information Technology, E\" otv\" os
          University, Budapest, P\'azm\'any P. s. 1/A, H-1117,
          Hungary\\
          \email{bagoly@ludens.elte.hu}
\and  Department of Physics, Bolyai  Military University,
Budapest,
Box-12, H-1456, Hungary\\
           \email{Horvath.Istvan@zmne.hu}
           \and Astronomical Institute of the Charles University, V Hole\v{s}ovi\v{c}k\'ach 2,
           180 00 Prague 8, Czech Republic\\
           \email{meszaros@cesnet.cz}
      }

\date{Received; accepted; published online}

\abstract{We studied the statistical properties of the luminosity
distance and internal extinction data of type Ia supernovae in the
lists published by Tonry et al. (\cite{Tonry03}) and Barris et al.
(\cite{Barris04}). After selecting the luminosity distance in an
empty Universe as a reference level we divided the sample into low
$z<0.25$ and high $z \ge 0.25$ parts. We further divided these
subsamples by the median of the internal extinction. Performing
sign tests using the standardized residuals between the estimated
logarithmic luminosity distances  and those of an empty universe,
on the four subsamples separately, we recognized that the
residuals were distributed symmetrically in the low redshift
region, independently from the internal extinction. On the
contrary, the low extinction part of the data of $z \ge 0.25$
clearly showed an excess of the points with respect to an empty
Universe which was not the case in the high extinction region.
This diversity pointed to an interrelation between the estimated
luminosity distance and internal extinction. To characterize
quantitatively this interrelation we introduced a hidden variable
making use of the technics of factor analysis. After subtracting
that part of the residual which was explained by the hidden
variable we obtained luminosity distances which were already free
from interrelation with internal extinction. Fitting the corrected
luminosity distances with cosmological models we concluded that
the SN Ia data alone did not exclude the possibility of the
$\Lambda=0$ solution. }


\authorrunning{Bal\'azs et al.}
\titlerunning{A possible interrelation between the luminosity
distances and internal extinctions of SNae Ia}

\maketitle

\section{Introduction}

The Ia type supernovae are unique tools for studying several
important large scale properties of the Universe. In particular,
their role in proving the existence of the positive value of the
cosmological constant appeared to be fundamental. There are
several coordinated efforts to get a large number of SN Ia of high
z to give a statistically firm support for the nonzero
cosmological constant (e.g. Riess et al. \cite{Riess98},
\cite{Riess04}; Perlmutter et al. \cite{Perl99}; Tonry et al.
\cite{Tonry03}; Barris et al. \cite{Barris04}; Astier et al.
\cite{Astier06}; but see also Gott et al. \cite{Gott01};
M\'esz\'aros \cite{Mesz02}; Rowan-Robinson \cite{RRob02}). A
comprehensive review on the cosmological implications from
observations of type Ia supernovae was published by Leibundgut
(\cite{Leib01}).

In reducing the raw observational data a crucial point is the
estimation of the effect of the interstellar and intergalactic
dust. All kinds of extinction have to be removed from the data
before using it for testing cosmological models. The foreground
extinction of our Milky Way is well studied and can  be removed
with certainty (Schlegel et al. \cite{Schlegel98}). Several
studies indicate that the effect of the intergalactic dust does
not play a significant role as well (Leibundgut \cite{Leib01} and
the references therein). The presence of the intergalactic dust
would increase the mean color excess of the distant objects.
Unless this dust distributes very homogeneously its effect would
increase also the scatter of the observed colors and magnitudes of
the distant supernovae in a clear contradiction to what is
observed (Perlmutter et al. \cite{Perl99}). The second property is
also valid for the color-independent (grey) intergalactic
extinction. Estimating the intrinsic extinction of the host galaxy
poses the most serious problem.

The observed extinction originating in the host galaxy depends not
only on the amount and properties of the dust content but also on
the location of supernovae within the host. The morphological type
of host galaxies may vary on a wide scale from the early type
ellipticals to the late type spirals and irregulars indicating a
wide range of dust content. There are controversial results on the
average properties of the dust in the SN Ia host galaxies of high
redshift. Estimating the  B-V, V-R, and V-I color excess for 20 SN
Ia's Riess et al. (\cite{Riess96}) concluded that the ratios of
selective to total extinction from dust in distant galaxies
hosting SN Ia's are consistent with the galactic extinction law.
In a recent paper Knop et al. (\cite{Knop03}) did not find
evidence for anomalous reddening from an independent set of 11
high-redshift supernovae observed with the Hubble Space Telescope.

In contrast, Clements et al (\cite{Clement04}) presented deep
sub-millimetre observations of sixteen galaxies at z=0.5, selected
through being hosts of a type Ia supernova and suggested that dust
in supernova host galaxies at z=0.5 could produce a dimming that
is comparable to the dimming attributed to accelerated expansion
of the Universe. Based on deep submillimeter observations of 17
galaxies at z=0.5 that are hosts of a type Ia supernova Farrah et
al. (\cite{Farrah04}) emphasized the need to carefully monitor
dust extinction when using type Ia supernovae to measure the
cosmological parameters. Reindl et al. (\cite{Reindl05}) found
that the law of interstellar extinction in the path length to the
SN in the host galaxy is different from the local Galactic law.

All these controversial results indicate that our knowledge on the
properties of dust in SN Ia host galaxies is far from being
complete. This situation motivated us to study thoroughly the
statistical properties of the existing SN Ia data in particular an
eventual systematical error in the available luminosity distance
and host galaxy extinction estimations.

This paper is organized as follows: Section 2 summarizes the
statistical properties of the data and claims to find an
interrelation between the estimated luminosity distance and
internal extinction. Section 3 deals with testing cosmological
models on the corrected data. Section 4 summarizes the main
results of the paper. The main points of this work were summarized
previously in the paper of Hetesi and Bal\'azs (\cite{Hetesi05}).

\section{Statistical properties of the data} \label{stat}

\subsection{Descriptive statistics of the sample} \label{desc}

The most comprehensive list of SN Ia objects  is the compilation
published by Tonry et al. (\cite{Tonry03}). It consists of 230 SN
Ia events but out of them only 188 have given internal host
extinction. We added to this sample  23 more SN Ia of high $z$,
published by Barris et al. (\cite{Barris04}). The mean error of
the estimated logarithmic luminosity distances amounted to 0.04
with a standard deviation of 0.02. After rejecting 10 outliers,
having uncertainties exceeding more than $3\sigma$ of the mean
standard deviation of the estimated luminosity distances, we got a
sample size of 201 objects for the further statistical
analysis\footnote{The compilation of Riess et al. (\cite{Riess04})
consists of 186 SNIa objects, having a significant overlap with
those of Tonty et al. (\cite{Tonry03}). Based on this sample Jain
and Ralston (\cite{Jain05}) pointed out a correlation between the
residuals of the luminosity distances and the estimated extinction
of the hosts. They did not attempt, however, to remove the effect
of this interrelation from the distances.}. The standard deviation
of the errors of the distances did not depend on the value of
extinction. Using this fact we gave to all extinction data equal
weight in the further analysis although their error was not given
in the list of Tonry et al. (\cite{Tonry03}). The estimated host
extinctions spread over a region of $A_V=0-4.1$ magnitude but
except of a few outliers the bulk majority of data are
concentrated in the 0-1 magnitude range.

\begin{figure}
  {\includegraphics[width=8.0cm]{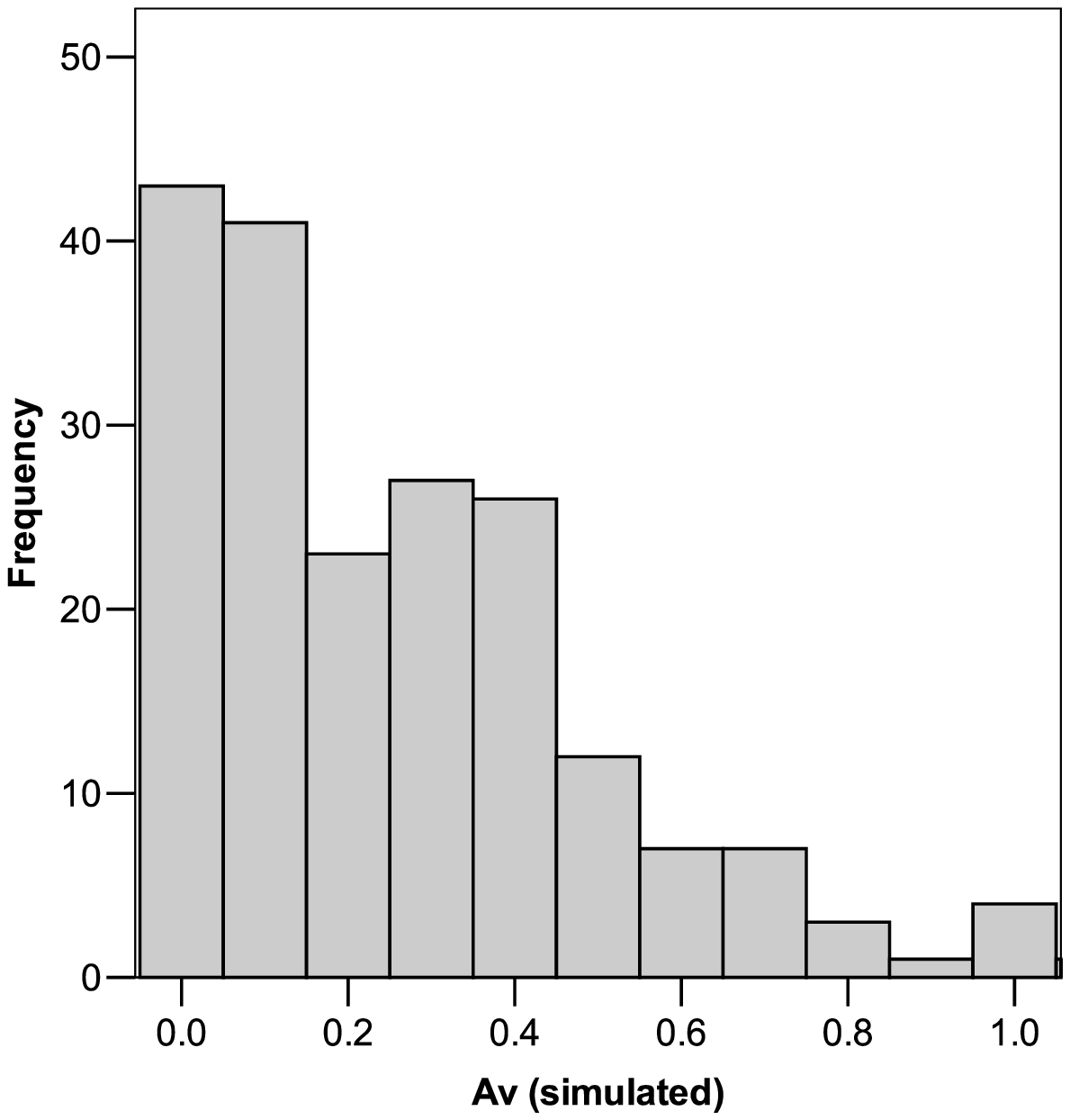}}
  {\includegraphics[width=8.0cm]{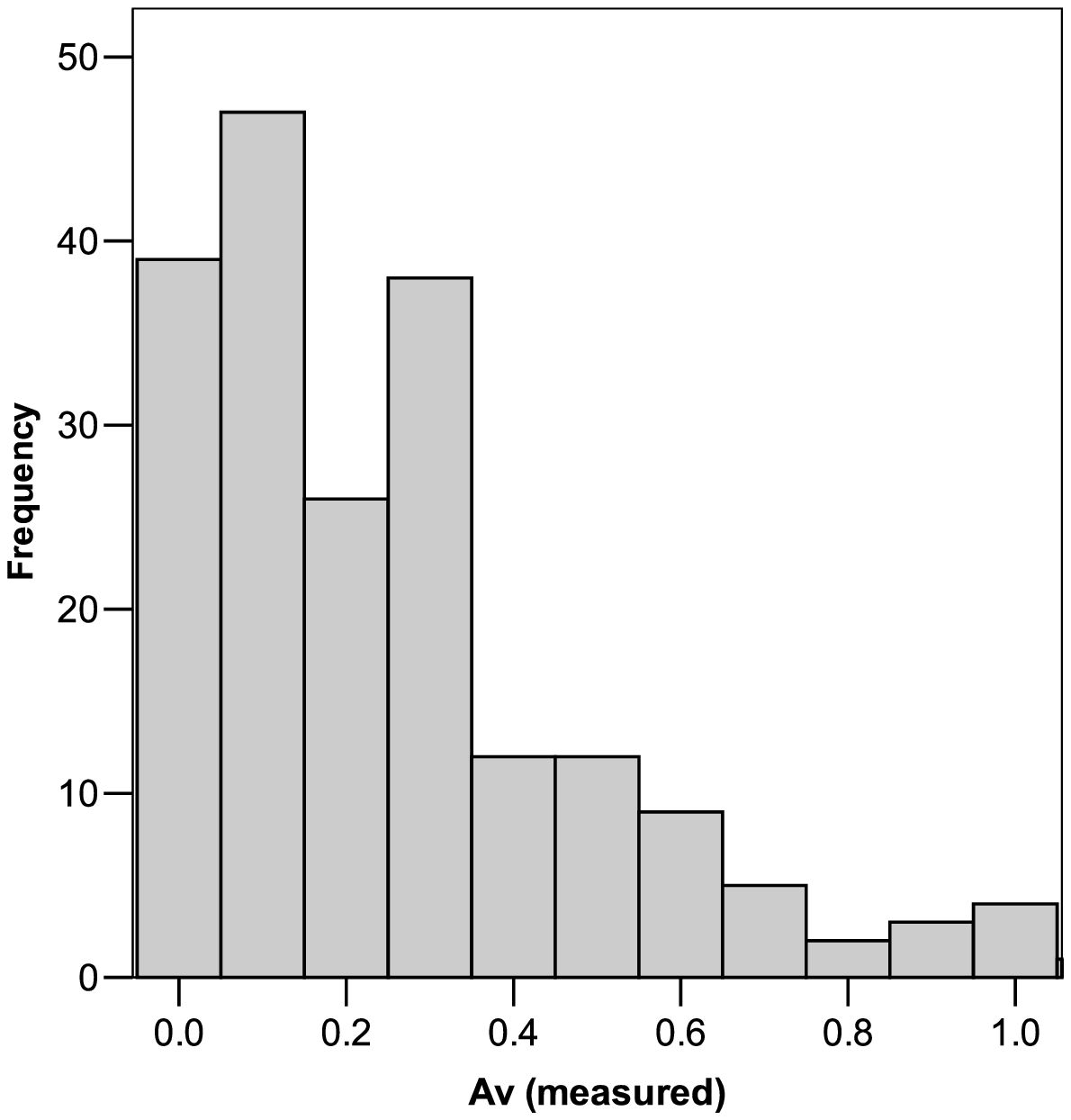}}
\caption{Histogram of simulated SNe extinctions (top), and the
real sample
   (bottom). The extinction is in magnitude on the horizontal axis. There  is a well pronounced
   peak at $A_V=0$ due to the objects at the front side of the host galaxy.
   The second smaller peak can be accounted for objects at the opposite side.
    According to a
   Kolmogorov-Smirnov test  the difference between  the simulated
   and  measured distributions can be accounted for chance with a $p=0.547$
   probability.}
\label{simabs}
\end{figure}

The distribution of the internal extinction can be easily modelled
by random orientation of a dusty galactic disc to the line of
sight\footnote{This is of course a very crude approximation since
the morphological types of the SN Ia host galaxies show a wide
variety of morphologies, including undisturbed ellipticals,
spirals and disturbed systems (Farrah et al. \cite{Farrah02}).}.
Assuming that the SN Ia supernovae belong to the old disc
population with a scale height of 300 pc (della Valle \& Panagia
\cite{Valle92}; Wainscoat et al. \cite{Wain92} gave 325 pc for the
old disc population) and the dust layer in the disc have a scale
height of about 130 pc and 0.18 mag visual extinction
perpendicular to the symmetry plane one gets the simulated
distribution as displayed in the upper panel of Figure
\ref{simabs}. The lower panel of Figure \ref{simabs} shows the
distribution of observed extinctions for comparison. The
pronounced peak at the low extinction part of the histogram is
probable accounted for supernovae in front side of the host galaxy
while the second less pronounced one is caused by those at the
opposite side. The observed distribution is probable biased at the
high redshift part of the sample towards lower extinctions because
close to the limit of detection the heavily obscured supernovae
can be missed by the observation. Despite of the simplicity of the
model applied the simulated data reproduce remarkably well the
basic characteristics of the observed distribution within the
limits of statistical inference. A Kolmogorov-Smirnov test
revealed that the difference between  the simulated and  measured
distributions can be accounted for chance with a $p=0.547$
probability.

\begin{figure}
  {\includegraphics[width=8.0cm]{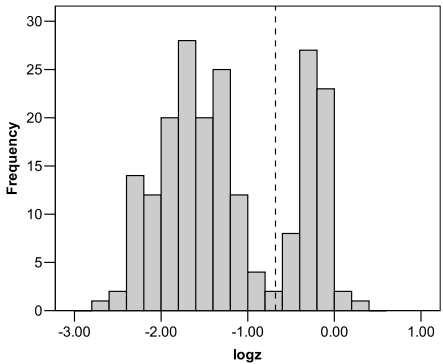}}
\caption{Histogram of z distribution in the data set. The vertical
   dashed line indicates a cut between the low and high redshift part of the sample.}
\label{zhist}
\end{figure}

The redshift distribution of the sample is bimodal (see Fig.
\ref{zhist}). There is a dip at z = 0.25 splitting the data set
into a low and a high redshift part. In  the low redshift domain
the luminosity distances derived from various cosmological models
differ only within the statistical uncertainty of estimating them
directly from SN Ia events. At the same time, the accuracy of the
measurements allows us to distinguish between cosmological models
in the $z> 0.25$ range. Assuming that $z$ is given, the SN Ia
luminosity distance, derived from an empty Universe, represents an
upper bound among those obtained from $\Lambda=0$ cosmological
models. In what follows, we consider this bound as a reference
distance at a given $z$. If the distances obtained from SN Ia
observation exceeded significantly this bound the introduction of
a nonzero cosmological constant seems to be inevitable.

Tonry et al. (\cite{Tonry03}) multiplied   the luminosity
distances by the present value of the $H_0$ Hubble constant and
listed their logarithm, along with the estimated uncertainty.

For further statistical studies we calculated the $s=(ld-ld_0)/
\sigma_l$ standardized deviation from an empty model, where $ld$
is the measured logarithmic luminosity distance  as given in Tonry
et al. (\cite{Tonry03}), $\sigma_l$ the corresponding measurement
error and $ld_0$ the calculated logarithmic distance in a given
model universe.

For given values of $\Omega_{\Lambda}, \Omega_{Matter}$ distances
are calculated by the following equality (Carroll et al.
\cite{Carroll92}):
$$ d_l = \frac{c(1+z)}{H_0 \sqrt{|\Omega_k|}}\, S \bigl(
\sqrt{|\Omega_k |}  $$
\begin{equation}
\times \int \limits_0^z [(1+z')^2(1+\Omega_M z') -
z'(2+z')\Omega_\Lambda]^{-1/2} dz' \bigr)
\end{equation}

\noindent where $\Omega_\Lambda + \Omega_M + \Omega_k = 1$ and
$S(x)$ is defined as $\sinh(x),\, x$, and $\sin(x)$ for $\Omega_k
>0,\, \Omega_k=0,$ and $\Omega_k <0$, respectively.

The $s$ deviations of the measured distances from the calculated
ones, based on some cosmological model, distributed symmetrically
in the case of a good fit. In the following we performed several
sign tests for the whole sample and several subsamples splitting
the original one with  $z=0.25$ and the median of the $A_V$
extinction. We use the median instead of the sample mean value
because the former is less sensitive to outliers. Table \ref{amed}
shows these quantities for the two redshift range defined and the
whole sample, respectively.

\begin{table}[h]
\caption{Statistical mean and median of the $A_V$ internal extinction
  for the low and high $z$ part and for the whole sample. Note the difference
  in the values between the low and high redshift part of the sample.
  The difference may be accounted for the observing bias in the high $z$
  part.}
\label{amed}
\begin{tabular}{lccr}\hline \hline
redshift   & Mean    & Median &  N \\
\hline
$z<0.25$     & 0.306 &   0.250 & 140  \\
$z \ge 0.25$ & 0.194  &  0.140 &  61 \\
\hline
Total        & 0.272  &  0.190 & 201 \\
\end{tabular}
\end{table}

\subsection{Performing  tests on the data} \label{signt}

A combination of the median of $A_V$ internal
extinction and the $z=0.25$ cut, divided the data set into four
subsamples. We used sign tests to study whether the distribution
of the standardized residuals, introduced in the \ref{desc}
subsection, was symmetric in relation to the level defined by the
empty model, within these subsamples. A significant excess of the
"+" signs favors a $\Lambda \neq 0$ model and in the opposite case
one gets support for traditional Friedmann solutions.

We made the assumption that the probability of the "+" (or "-")
signs of the $s$ residuals followed a Binomial distribution in the
form of:

\begin{equation}
P(k) = {n \choose k} p^k (1-p)^{n-k} \label{binom}
\end{equation}

\noindent where $n$ means the number of trials, $k$ that of
successes and $p$ is the probability of one of these events
($p=0.5$ in symmetric case). The mean value of successes is given
by $np$. One may compute the probability of the $|np-k| \le
\delta$ deviation from the mean assuming that it happens only by
chance:

\begin{equation}
 P(|np-k|\le \delta) = \sum \limits_{|np-k| \le \, \delta\, =\, |np-l|}
P_l \label{binprob}
\end{equation}

\begin{figure}
  {\includegraphics[width=8.0cm]{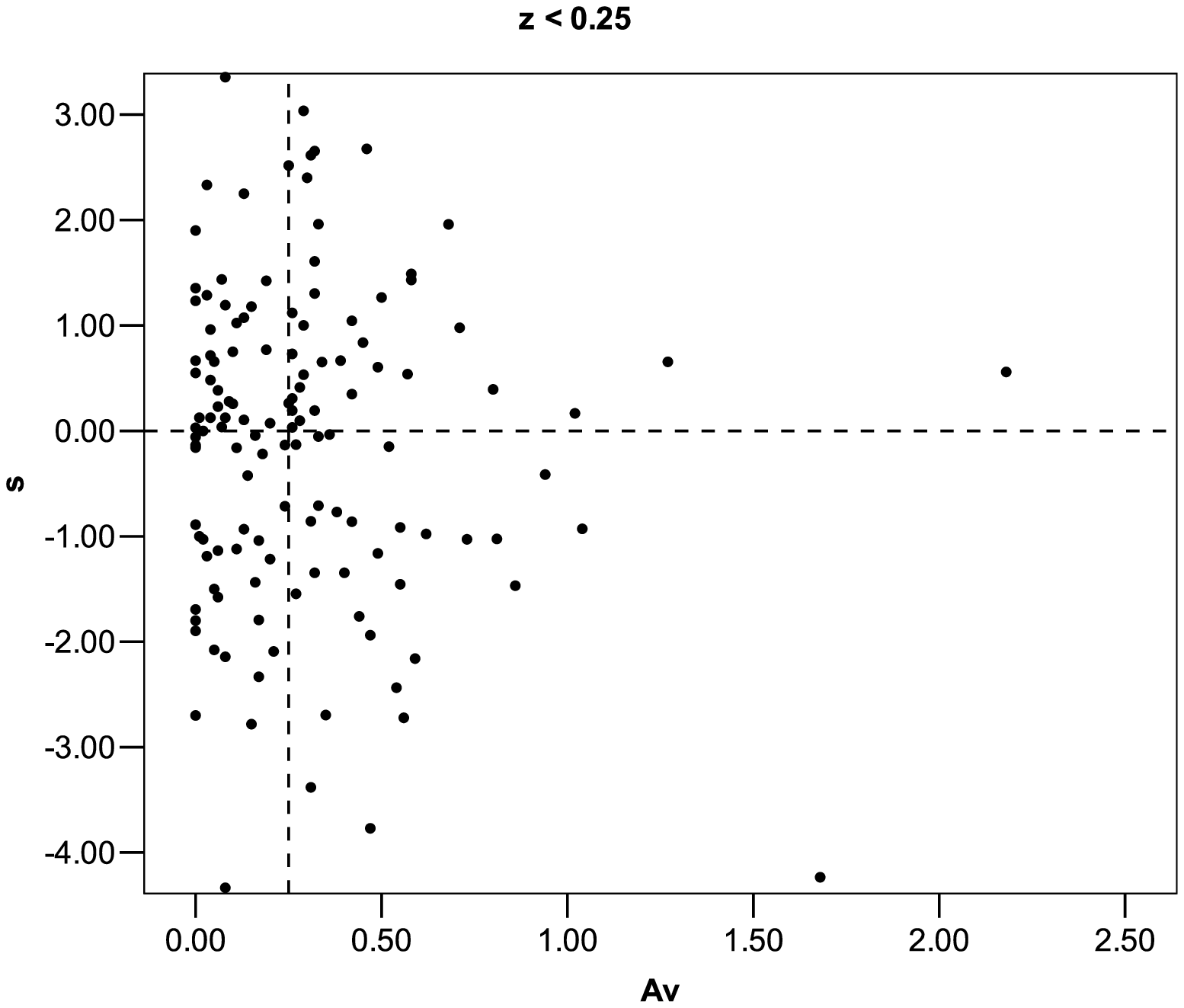}}
  {\includegraphics[width=8.0cm]{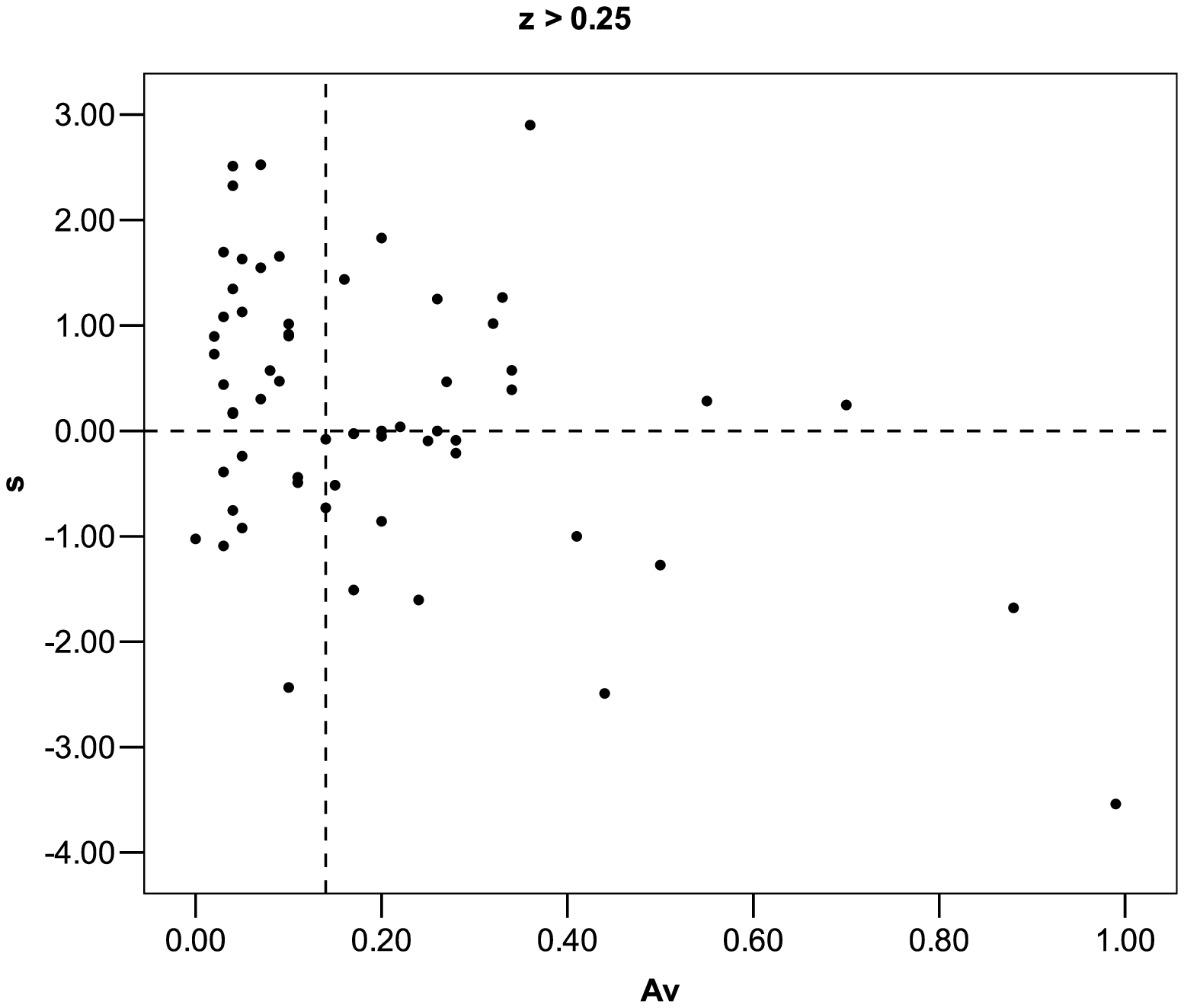}}
\caption{The extinction -- standardized residual relation of the
low (top)
   and high-$z$ (bottom) subset. Vertical dashed lines mark the median of the
   extinction data. Horizontal dashed lines mark the reference level of an
   empty Universe. Note the difference between the left and the right panel.
   In the $z < 0.25$ case the distribution of residuals is symmetric to the
   reference level of an empty Universe, independently from the extinction.
   On the contrary, the low extinction part (left from the median line) of
   the $z > 0.25$ panel clearly has an excess of the points above the
   reference line but it disappears at higher extinction values displaying
   a pronounced negative trend.}
\label{3}
\end{figure}

The formula given by Equation (\ref{binprob}) has the advantage
that it is exact and works correctly even in the case of a small
sample size. We carried out sign tests on the subsamples defined
by cutting the data set at the median in the low and high redshift
part, separately. Tables \ref{lowz} and \ref{highz} summarize the
results. As Table~\ref{lowz} demonstrates the subsamples are
symmetric, independently of the internal extinction. On the
contrary, the results summarized in Table~\ref{highz} indicates a
difference in the statistical properties of the low and high
extinction part of the subsample (see Figure \ref{3}). While  the
low extinction part clearly shows an excess of the points above
the reference line it is not the case among the points above the
$A_V$ median. This fact indicates some sort of interrelation
between the internal extinction and the standardized residuals in
the high $z$ domain.

\begin{table}[h]
\caption{Sign tests for the low redshift ($z < 0.25$)
part of the sample, divided by the median value of the $A_V$
internal extinction. The last column ($P_l$) gives the probability
that the numbers of "-" and "+" signs differ only by chance. The
distribution of the standardized residuals are symmetric,
independently on the value of $A_V$.} \label{lowz}
\begin{tabular}{ccccc}
\hline \hline
               & k ("-") & n-k ("+")& n   & $P_l$ \\
\hline
$A_V < 0.25$   & 34      & 35       & 69  & 1.000 \\
$A_V \ge 0.25$ & 34      & 37       & 71  & 0.813 \\
\hline Total   & 68      & 72       & 140 & 0.800 \\
\hline
\end{tabular}
\end{table}

\begin{table}[h]
\caption{Sign tests for the high redshift ($z \ge
0.25$) part of the sample, divided by the median value of the
$A_V$ internal extinction. The last column ($P_l$) gives the
probability that the numbers of "-" and "+" signs differ only by
chance. Unlike to Table \ref{lowz} the low extinction part of the
sample has an excess of the "+" signs which can not be accounted
only by chance. On the contrary, the distribution in the $A_V \ge
0.140$ domain is still symmetric. The difference in the behavior
between the $A_V < 0.140$ and $A_V \ge 0.140$ part of the sample
indicates an interrelation between the $s$ normalized residuals
and the $A_V$ internal extinction.} \label{highz}
\begin{tabular}{ccccc}
\hline \hline
                & k ("-") & n-k ("+")& n  & $P_l$ \\
\hline
$A_V < 0.140$   & 9       & 21       & 30 & 0.043 \\
$A_V \ge 0.140$ & 17      & 14       & 31 & 0.720 \\
\hline Total    & 26      & 35       & 61 & 0.305 \\
\hline
\end{tabular}
\end{table}

The sign test is appropriate for studying the significance of the
excess of points above the $s=0$ line and could give a clear
support for $\Lambda\neq 0$ cosmology. One may wonder, however,
whether additional statistical tests also confirm the result
obtained from the sign test. To proceed in this way we performed
three additional tests on the high $z$ part of the sample:
Student's $t$, Mann-Whittney and Kolmogorov-Smirnov tests. The
Student's test compares the means while the others the
distribution of $s$ at  both side of the median of $A_V$. We
summarized the results in Table \ref{tests}.

\begin{table}[h]
\caption{Additional tests on the high $z$ part of the sample}
\label{tests} \begin{tabular}{lcc} \hline \hline Type of test &
sign. & No. of objects\\ \hline
Student's $t$ & 0.044 & 61\\
Mann-Whittney & 0.025 & 61\\
Kolmogorov-Smirnov & 0.005 & 61 \\
\hline
\end{tabular}
\end{table}

 One may infer from Table \ref{tests} that the results of the additional tests
 confirm the conclusion obtained from the sign test.

\subsection{Pearson's correlation and factor analysis}

To quantify the interrelation pointed out in subsection
\ref{signt} we computed Pearson's linear correlation between $A_V$
and $s$ in the low and high $z$ domain, separately. Inspecting
Table \ref{pearson} infers that there is a very significant
correlation in the high redshift part unlike the low redshift one.

We made the assumption there was an $f$ hidden variable which was
present in both $s$ and $A_V$ and it was responsible for the
interrelation. From strictly statistical point of view, it was not
necessary to specify the physical nature of this variable.
Following this assumption, the observed $s$ and $A_V$ could be
expressed in terms of $f$ by the following system of equations:


\begin{equation} \label{As}
\pmatrix {A_V  \cr s \cr} =  \pmatrix {A_0  \cr s_0 \cr} f +
\pmatrix {\varepsilon_A  \cr \varepsilon_s \cr}
\end{equation}

\noindent where $A_0$, $s_0$ are constants and $\varepsilon_A$,
$\varepsilon_s$ represent noise terms. To estimate $A_0$, $s_0$
and $f$ we invoked the standard technique of factor analysis. In
many statistical software packages (e.g. in the widely used
SPSS\footnote{SPSS is a registered trade mark for Statistical
Package for Social Sciences}) the default solution of a factor
model is equivalent to the computation of principal components
(PC) of the correlation matrix of the observed variables ($A_V$
and $s$ in our case). For obtaining PCs one has to solve the
eigenvalue problem of the correlation matrix of the observed
variables:

\begin{equation}
   \pmatrix {1 & r \cr r & 1 \cr} \pmatrix {a_1  \cr a_2 \cr} =
   \lambda \pmatrix {a_1 \cr a_2 \cr}
\end{equation}

\noindent where $r$ is the Pearson correlation between $A_V$ and
$s$, $\{a_1,a_2\}$ are the components of the eigenvector, and
$\lambda$ is the eigenvalue of the matrix.

If one used principal components for establishing  a factor model
the usual way is to make a distinction between the 'significant'
and 'non-significant' PCs and keeping only the former ones. A
common procedure for making this kind of distinction is to keep
only those eigenvectors which belong to eigenvalues above a given
threshold. The widely used Kaiser criterion (Kaiser
\cite{Kaiser60}) keeps eigenvalues of $\lambda\geq 1$. We used
this criterion in our analysis.

The $\{a_1,a_2\}$ components can be used to obtain the constants
in Equation (\ref{As}) according to the following equalities:

\begin{equation}
A_0=\sigma_{A_V} a_1 \phantom{@@@};  \phantom{@@@} s_0= \sigma_s
a_2 \phantom{@},
\end{equation}

\noindent where $\sigma_{A_V}; \sigma_s $ are the standard
deviations of $A_V$ and $s$, respectively. For each data point the
value of the $f$ variable (the {\it factor score}) is resulted
with keeping the first PC and $\varepsilon_A$ ; $\varepsilon_s$
are the residuals when substituted $f$ into Equation (\ref{As}).

\begin{table}[h]
\caption{Pearson's linear correlation between the
 internal extinction and the $s$
standardized residual. The third column gives the probability for
the purely random occurrence of the correlation.}
\label{pearson}
\begin{tabular}{cccc} \hline \hline
             & $CORR(A_V,s)$ & Sig. (2-tailed) & N \\
\hline
$z < 0.25$   & -0.068        & 0.427           & 140 \\
$z \ge 0.25$ & -0.411        & 0.001           & 61 \\
\hline Total & -0.122        & 0.086           & 201 \\
\hline
\end{tabular}
\end{table}

\begin{figure}
  {\includegraphics[width=8.0cm]{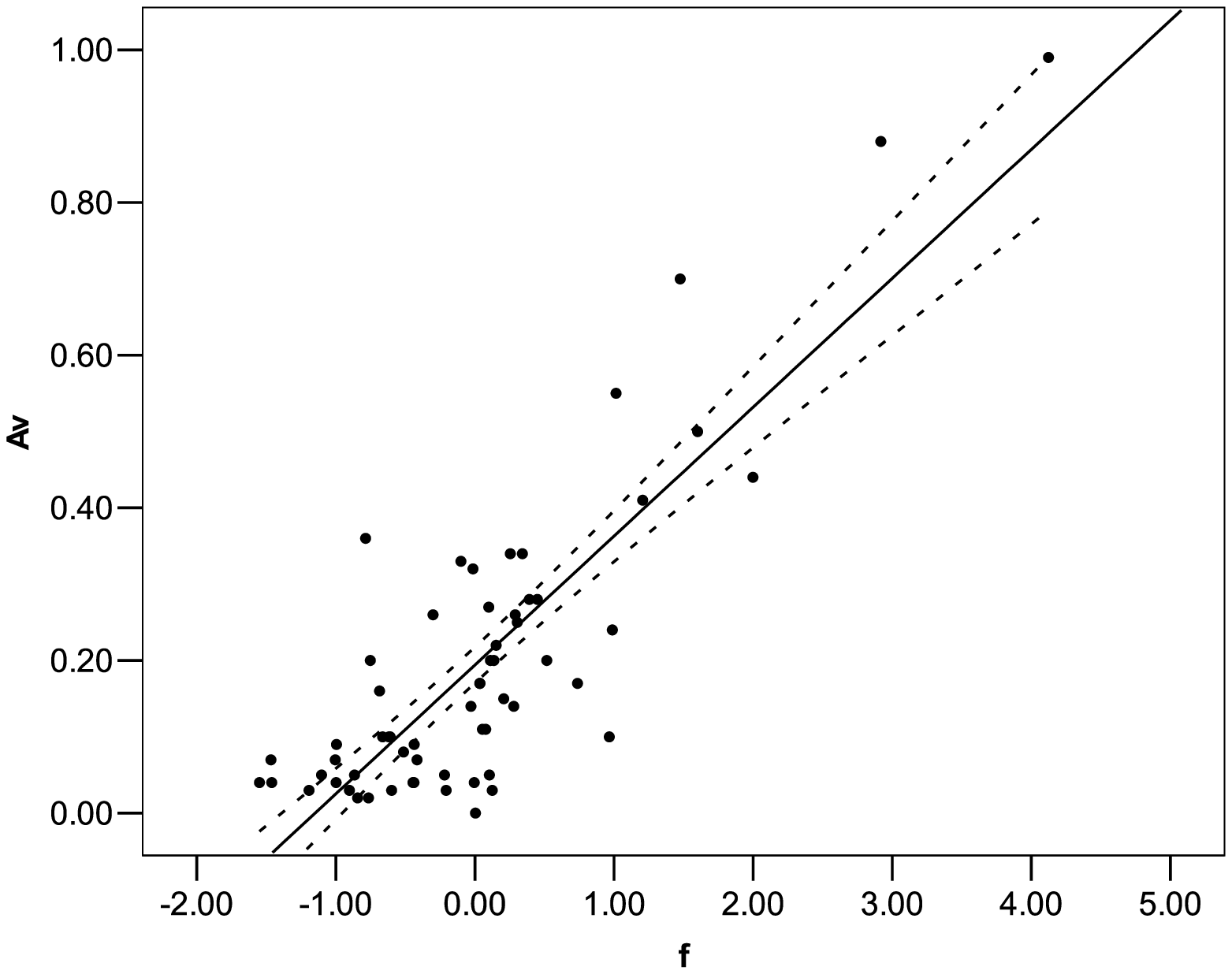}}
  {\includegraphics[width=8.0cm]{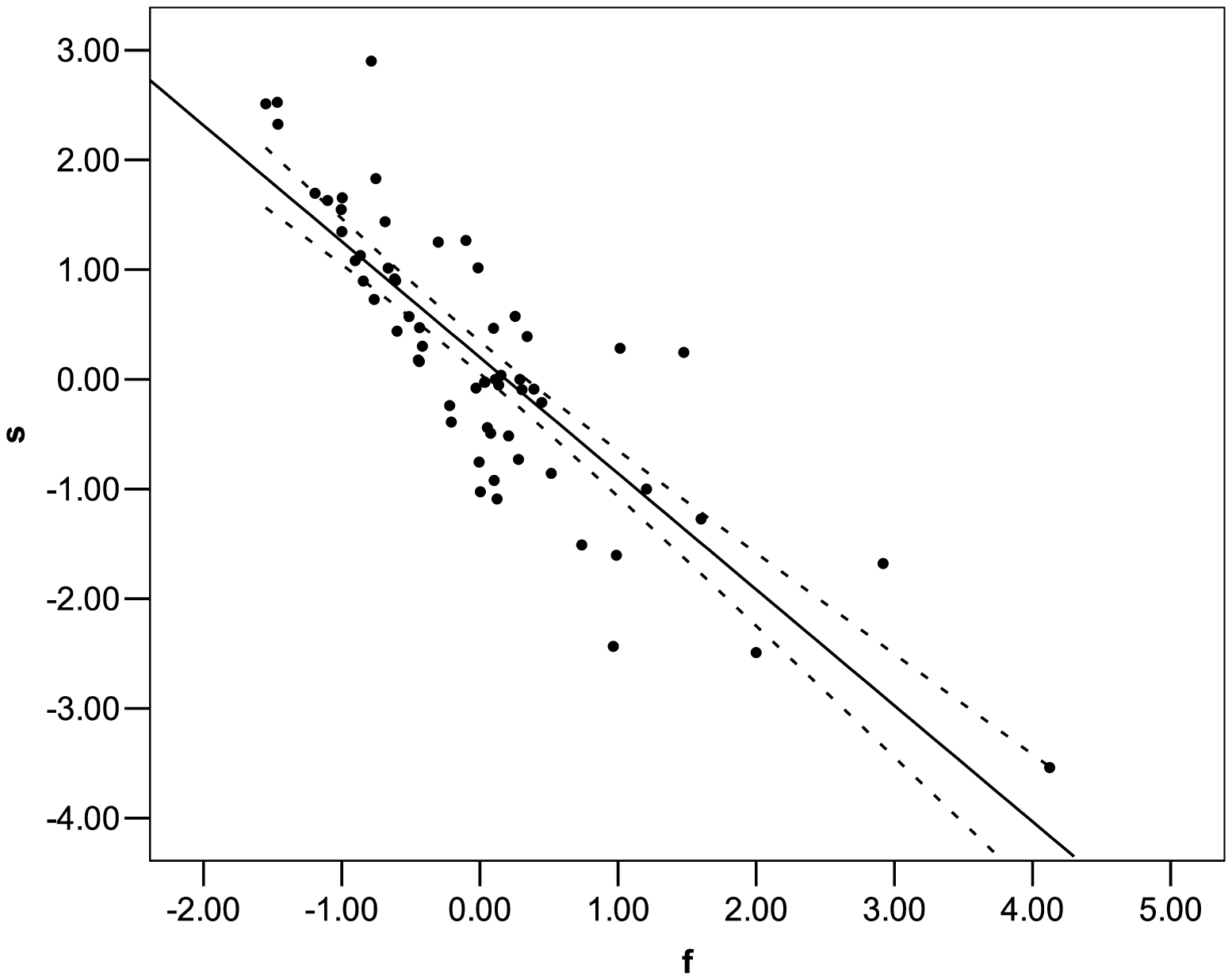}}
\caption{Dependence of the $A_V$ internal extinction (upper panel)
and the $s$ residual (lower panel) on the $f$ background variable
obtained from the factor analysis. Full lines mark the
relationships between the observed and hidden variables as
obtained from the factor analysis and dashed ones define the 95\%
confidence region for the fit. The effect of the hidden variable
has to be removed from the luminosity distances before using them
for testing cosmological models.} \label{pc}
\end{figure}

The $f$ background variable reproduced well both the $A_V$
internal extinction and the $s$ residual as displayed in Figure
\ref{pc}. One has to  remove the effect of this variable from the
luminosity distances  before using them for testing cosmological
models.

\subsubsection{Caveats}

The high level of significance obtained for the Pearson's
correlation has to be treated with some caution. Usually, one
assumes that both variables have Gaussian distribution when
calculating the significance level of the Pearson's correlation.
In our case this assumption is far from the reality in particular
at $A_V$ in the $z > 0.25$ range. There are 57 objects in the $0 <
A_V \le 0.5$ part and only 4 with $0.5 < A_V < 1.0$.

If one considered only the data points  in the $0<A_V\le 0.5$
range of the $z > 0.25$ sample the Pearson's correlation is still
negative but drops back to -0.221 with a significance of 0.094.

To get a measure of correlation independent on the assumption  of
normality we split the sample into four parts with the medians of
$s$ and $A_V$. The number of data points in these quadrants form a
$2\times 2$ contingency table.  Making use this table one can
compute other measures of correlation as listed in Table
\ref{corr}.

\begin{table}[h]
\caption{Different types of correlation in the high redshift part
of the sample} \label{corr}
\begin{tabular}{lccc} \hline \hline
   Type of corr. & $CORR $ & Sig.  & N \\
\hline
Pearson's linear ($0<A_V<1.0)$   & -0.411  & 0.001      & 61 \\
Pearson's linear ($0<A_V \le 0.5$) & -0.221 & 0.094      & 57 \\
Pearson's R ($2\times2$ cont. table) & -0.381 & 0.005   & 57 \\
Spearman ($2\times2$ cont. table) & -0.381        & 0.005 & 57 \\
\hline
\end{tabular}
\end{table}

One may infer from Table \ref{corr} that  measure of correlation
obtained from the contingency table is close to the Pearson's
linear correlation computed for the whole $z> 0.25$ sample.

\section{Testing cosmological models}

As we pointed out at the end of Section \ref{stat} the luminosity
distances had to be freed from the interrelation with the internal
extinction, before testing cosmological models. The factor
analysis splits the actual residual into two parts: $s=s'+s"$
where $s'$ represents that part explained by the $f$ background
variable while $s"$ stands for the  true one which is independent
from the amount of the internal extinction.

Going back to the definition of $s$ as given in Subsection
\ref{desc} we may write $ld=ld_0+\sigma_l (s'+s")$. Subtracting
$s'$ from this equation, that part explained by the background
variable,  we obtain: $ld^c = ld - \sigma_l s' = ld_0+\sigma_l s"$
which is already free from the interrelation with internal
extinction. This luminosity distance can be used for testing
cosmological models.

The luminosity distance of an object of $z$ redshift can be
calculated in the framework of a given cosmological model. Besides
$z$ it depends on $\Omega_\Lambda$ and $\Omega_M$ cosmological
parameters. It is widely accepted, based on the recent results of
SN~Ia data and the WMAP project, that $\Omega_\Lambda=0.7$ and
$\Omega_M=0.3$ favoring a $\Omega_\Lambda+\Omega_M=1$ flat
Euclidean model (Hansen et al. \cite{Hansen04}; Verde
\cite{Verde03}; Wright \cite{Wright03}).

We did not intend to repeat the estimation of $\Omega_\Lambda$ and
$\Omega_M$ using the original data of Tonry et al.
(\cite{Tonry03}) and Barris et al. (\cite{Barris04}). To make a
comparison we used only our corrected data. For this purpose we
calculated the $\chi^2$ distance between our corrected data and
those obtained from cosmological models according to the equation
below:

\begin{equation} \label{khi}
\chi^2=\sum \limits_{i=1}^n
\frac{(ld^c-ld^{calc}(z,\Omega_\Lambda, \Omega_M))^2}{\sigma_l^2}
\end{equation}

\noindent Minimizing this distance with respect to
$\Omega_\Lambda, \Omega_M$ gives the best estimate of these
parameters according to the observed set of $ld^c$.  Assuming that
the $ld^c-ld^{calc}$ residuals follow a Gaussian distribution and
introducing the $L=\log P$ likelihood function we obtain:

\begin{equation}\label{L}
L= -\frac{\chi^2}{2}+constant
\end{equation}

\noindent The confidence of the parameters can be estimated making
use of the following theorem :

\begin{equation}\label{lmax}
2(L_{max} - L_0) \approx \chi^2_p
\end{equation}

\noindent where $L_{max}, L_0$ mean  the likelihood function at
the maximum and the true value of the parameters estimated,
respectively (for the details see Kendall \& Stuart \cite{KS73}).
The degree of freedom  is $p=2$ in our case.

Based on our corrected luminosity distances we calculated the best
fitting values of $\Omega_\Lambda, \Omega_M$ by minimizing
$\chi^2$ in Equation (\ref{khi}) and, consequently,  maximizing
$L$ in Equation (\ref{L}) at the same time.

\begin{figure}
  {\includegraphics[width=8.0cm]{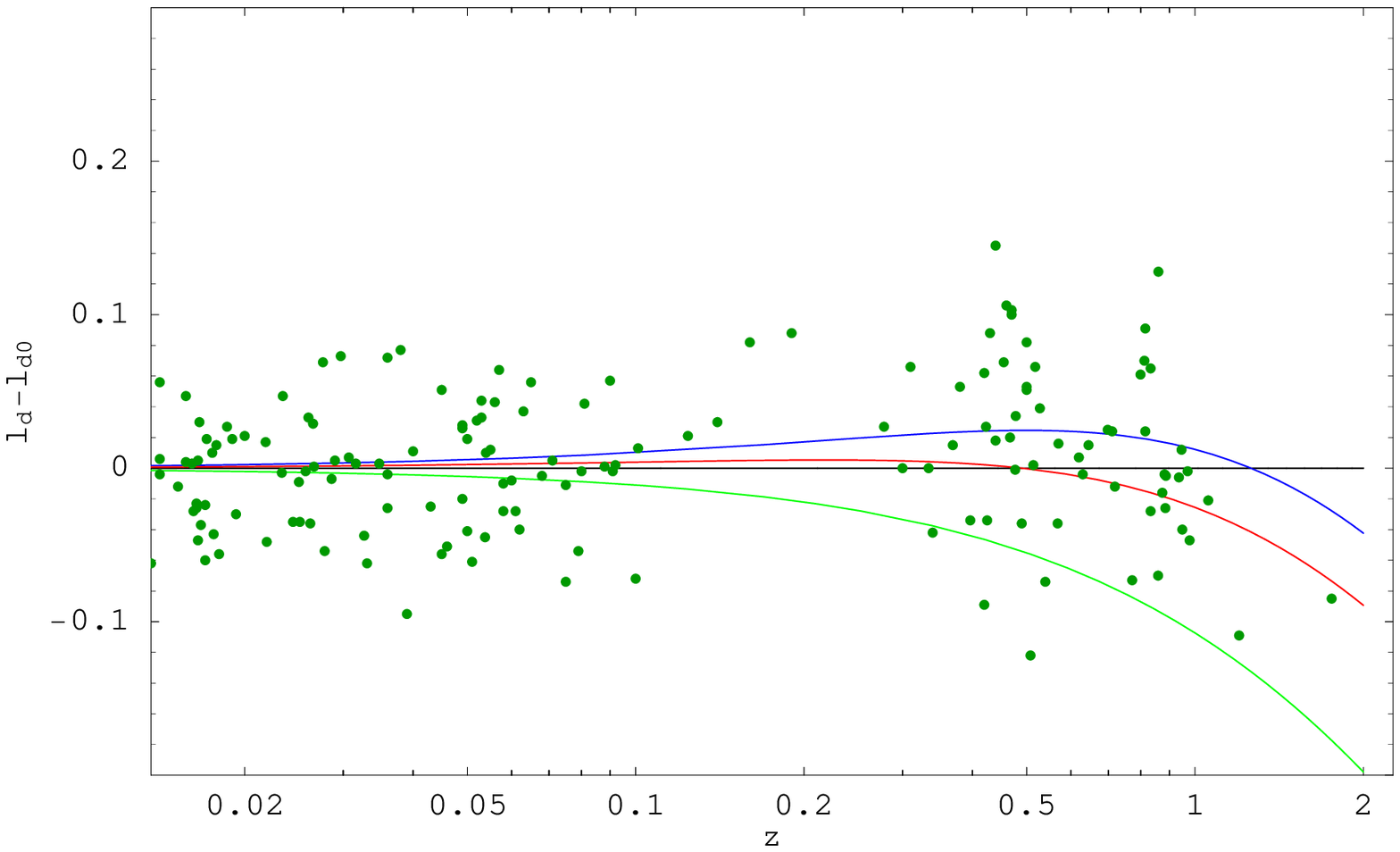}}
  {\includegraphics[width=8.0cm]{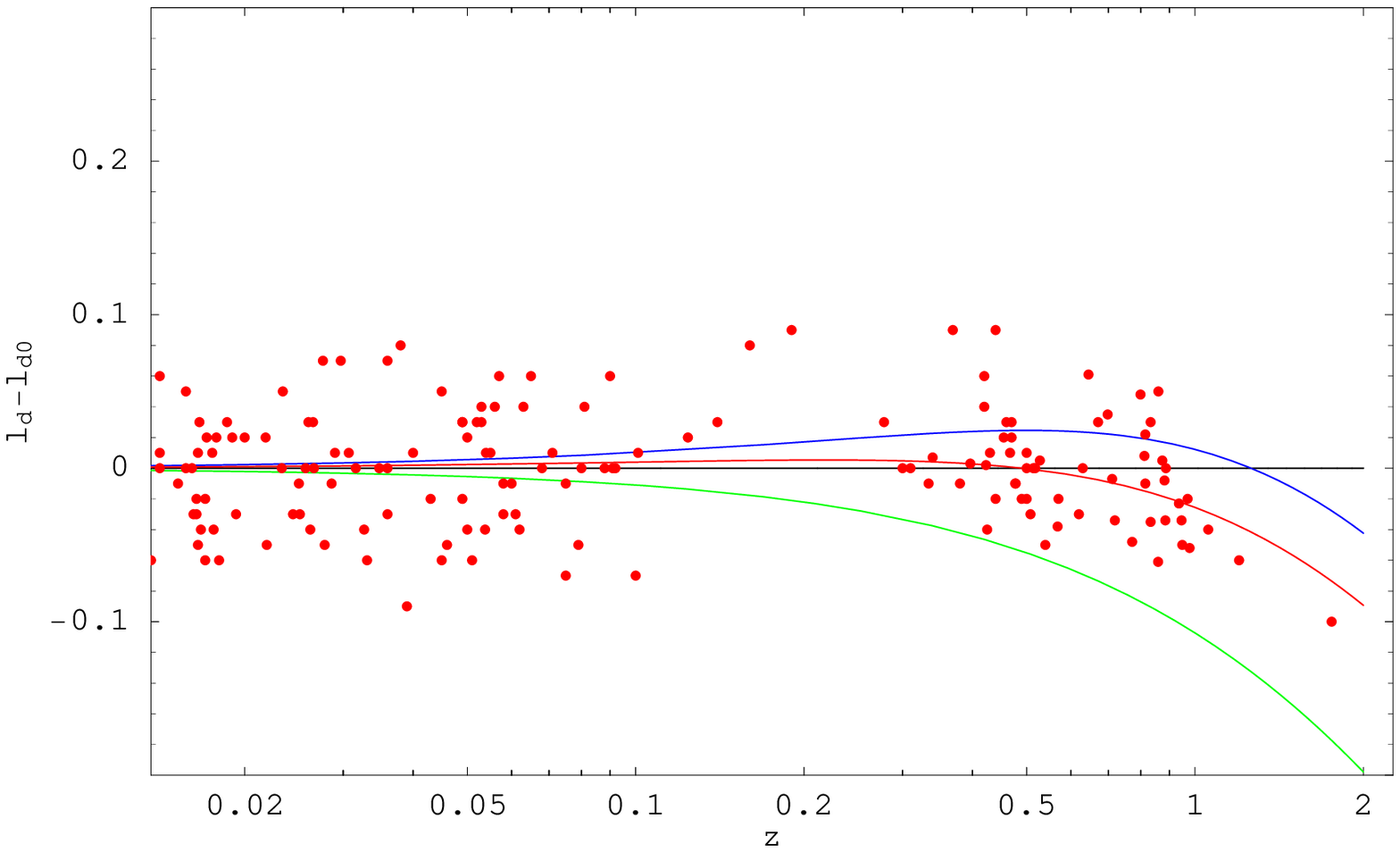}}
  \caption{Deviation of the observed logarithmic SN Ia luminosity
           distances from those in an empty Universe as a function of the $z$
           redshift. Upper panel shows this relationship in the case of
           uncorrected data and the lower one after correction. The best
           fitting solutions are also indicated. Note the much smaller
           scatter of the corrected data around the best fitting line in the
           $z > 0.25$ region in comparison with the non-corrected data.
           Blue
           and red lines mark the best fit to the corrected ($\Omega_\Lambda
           = 0.47$ ; $\Omega_M = 0.43 $) and
           the canonical  model ($\Omega_\Lambda
           = 0.7$ ; $\Omega_M = 0.3 $), respectively. Green line
            indicates the $\Lambda=0$ Euclidean model.}
\label{fits}
\end{figure}

 As a best fit we obtained $\Omega_\Lambda
= 0.47$ and $\Omega_M = 0.43 $ which considerably differ from the
canonical $\Omega_\Lambda =0.7$ and $\Omega_M = 0.3$. In Figure
\ref{fits} we displayed the scatterplot between $z$ and the
deviation of the observed luminosity distances and the calculated
ones in an empty Universe before (left panel) and after correction
(right panel). It is worth noticing the much smaller scatter of
the corrected data around the best fitting line in the $z > 0.25$
region in comparison with the non-corrected data. Computing the
$\chi^2$ value for one degree of freedom we obtained
$\chi^2_{df}=1.46$ for the non-corrected sample and
$\chi^2_{df}=0.38$ for the corrected one.

\begin{figure}[h]
 {\includegraphics[width=\linewidth]{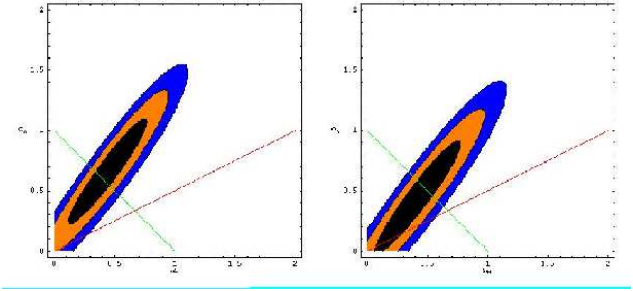}}
\caption{Confidence region of $\Omega_\Lambda, \Omega_M$ in case
         of the original data (left panel) and the same after correction
         (right panel). The outer boundaries of the black, orange and
         blue areas represent  67\%, 95\% and 99.3\% confidences,
         respectively. Red line separates the accelerating and decelerating and
         green line the open and closed
         model.
         Note that unlike the uncorrected case, the corrected
         data are consistent with the existence of a $\Lambda= 0$
         solution.}
\label{conf}
\end{figure}

Equation (\ref{lmax}) enabled us to calculate the confidence
interval of the parameters estimated. After fixing a value for
$\chi_p^2$ Equation (\ref{lmax}) specifies a boundary in the
$\{\Omega_\Lambda, \Omega_M\}$ parameter space. The particular
value specified for $\chi_p^2$ gives a $\beta$ probability that
the true value of  the estimated parameters are outside and
$1-\beta$ that inside this boundary. $1-\beta$ gives the
confidence level of this region. Following this procedure we
calculated the confidence and the result is shown in the right
panel of Figure \ref{conf}. In the left panel of this figure we
also give the confidence of the parameters using the uncorrected
sample for comparison.

A comparison of the left and the right  panel of Figure \ref{conf}
clearly shows the basic difference between  the parameters
obtained from the uncorrected and corrected data. While the
uncorrected data strongly supports the $\Lambda\neq 0$ solution
$\Lambda = 0$ is  well within the 95\% confidence region in the
corrected case. It is also important to note that the canonical
$\Omega_\Lambda =0.7$ and $\Omega_M = 0.3$ solution is outside of
the 99\% confidence region of the corrected data. However, the
$\Omega_\Lambda + \Omega_M = 1$ line referring to an Euclidean
Universe crosses the $1\sigma$ confidence region with a best
estimate of $\Omega_\Lambda =0.55$ and $\Omega_M = 0.45$ values.

The statistical results obtained in Section \ref{stat} are purely
phenomenological and give no hint of the probable reason of the
interrelation between the luminosity distance and internal
extinction as given in the data set of Tonry et al.
(\cite{Tonry03}) and Barris et al. (\cite{Barris04}). A
 detailed careful simulation of the procedure of obtaining the
 luminosity distance and internal extinction would be desirable to
 uncover the probable reason of the interrelation found in Section
 \ref{stat}. A detailed study of this kind may reveal whether
 this interrelation has some astrophysical reason or is purely a
 byproduct of the procedure for obtaining the luminosity distance
 and internal extinction.

\section{Summary and Conclusions}

We studied the basic statistical properties of the SN Ia sample
published by Tonry et al. (\cite{Tonry03}) and Barris et al.
(\cite{Barris04}). The observed distribution of the internal
extinction due to the host galaxies can be well modelled by a
dusty disc oriented randomly to the line of sight. We divided the
sample into the low and high $z$ part at a dip in the distribution
at $z=0.25$.

We further divided the low and the high redshift part of the
sample  by the median of the internal extinction. Four subsamples
resulted in this way. We selected the redshift - luminosity
distance relationship of an empty Universe as a reference level
and calculated the $s$ standardized deviation of the logarithmic
luminosity distances of Tonry et al. (\cite{Tonry03}) and Barris
et al. (\cite{Barris04}) from those of the empty model.

Performing sign tests on the standardized residuals, on the four
subsamples separately, revealed that $s$ is distributed
symmetrically in the low redshift ($z<0.25$) part of the sample,
independently on the internal extinction. On  the other hand, the
low extinction part of the data of $z \ge 0.25$ clearly showed  an
excess of the points of $s>0$ which was not the case among those
of having extinction above the median. This diversity in the
behavior of the points below and above the median pointed to an
interrelation between the $s$ residual and the $A_V$ internal
extinction.

To characterize quantitatively the interrelation between $s$ and
$A_V$ we computed the Pearson's linear correlation in the low and
the high redshift part, separately.  Assuming that the
interrelation can be represented by a hidden common variable in
$s$ and $A_V$ we constructed a factor model for explaining the
observed variables. We used PCA and the  Kaiser criterion for
verifying the factor model.

After subtracting that part of the residual which was explained by
the  factor model we obtained a corrected sample, free already
from the interrelation between $s$ and $A_V$. Minimizing the
$\chi^2$ distance between the corrected data and a cosmological
model with respect to $\Omega_\Lambda, \Omega_M$ one may conclude
that there was no need for $\Lambda\neq 0$.

We have to emphasize, however, than one came to this conclusion
assigning the same weight to the low and higher extinction part of
the sample in the $z > 0.25$ region. Nevertheless, some concerns
can be made for the reliability of the estimated extinction in the
host galaxy. Without the proposed correction the low extinction
part of the $z
> 0.25$ sample clearly supports the $\Lambda \neq 0$ cosmology.

Our result was purely phenomenological based only on the
statistical properties of the data. Consequently, further detailed
studies are required whether the interrelation between the
luminosity distance and the internal extinction in the high $z$
part of the sample has an astrophysical meaning or is simply a
byproduct  of the way as $ld$ and $A_V$ were obtained from the
data.

\acknowledgements{This work was supported by the OTKA grant
T048870. We are grateful to P.G. Teres for carefully reading the
manuscript and suggesting valuable improvements. Zs. Hetesi is
indebted to B. Bal\'azs for numerous discussions.}


\end{document}